%% file: main.tex
\documentclass[10pt]{article}

\input{math_commands.tex}

\usepackage{hyperref}
\usepackage{url}
\usepackage{todonotes}
\usepackage{nicefrac}
\usepackage{booktabs}
\usepackage{cleveref}
\usepackage{CJKutf8}
\usepackage{graphicx}
\usepackage{enumitem}
\usepackage{pifont}
\usepackage{natbib}
\usepackage{authblk}

\usepackage[margin=1in]{geometry}

\newcommand{\xmark}{\ding{55}}

\title{Enhancing Large Language Model-based Speech Recognition by Contextualization for Rare and Ambiguous Words}

\author[1]{Kento Nozawa}
\author[1]{Takashi Masuko}
\author[1]{Toru Taniguchi}

\affil[1]{Preferred Elements, Inc.}
\affil[ ]{\texttt {\{nzw0301, masuko, ttani\}@preferred.jp}}

\date{}

\begin{document}

\maketitle

\begin{abstract}
      We develop a large language model (LLM) based automatic speech recognition (ASR) system that can be contextualized by providing keywords as prior information in text prompts.
      We adopt decoder-only architecture and use our in-house LLM, PLaMo-100B,
      pre-trained from scratch using datasets dominated by Japanese and English texts as the decoder.
      We adopt a pre-trained Whisper encoder as an audio encoder, and the audio embeddings from the audio encoder are projected to the text embedding space by an adapter layer and concatenated with text embeddings converted from text prompts to form inputs to the decoder.
      By providing keywords as prior information in the text prompts, we can contextualize our LLM-based ASR system without modifying the model architecture to transcribe ambiguous words in the input audio accurately.
      Experimental results demonstrate that providing keywords to the decoder can significantly improve the recognition performance of rare and ambiguous words.
\end{abstract}

\section{Introduction}

In recent years, large language models (LLMs) have been attracting attention
thanks to their very high performance in natural language processing
\citep{Devlin2019NAACL,Brown2020NeurIPS,Llama-Team-AI-Meta2024techreport}.
The performance of LLMs can improve by increasing the amount of calculations,
the amount of data, and the number of model parameters,
as known as scaling law \citep{Kaplan2020arXiv}.
Such pre-trained LLMs can solve different downstream tasks
by contextualizing input texts, namely, prompts \citep{Brown2020NeurIPS}
without updating parameters.
In addition, supervised fine-tuning can further improve
the performance of LLMs for downstream tasks \citep{Wei2022ICLR}.
However, supervised fine-tuning requires
many computational resources to update the parameters of LLMs.
Often, parameter efficient fine-tuning \citep{Ding2022arXiv},
where the subset of parameters is fixed,
compromises the performance of downstream tasks and computational resources: the amount of memory.

There have also been many attempts to use voice input for decoder-only LLMs
to have the multi-modal capability \citep{Latif2023arXiv}.
In these models, outputs from audio encoders
are discretized \citep{DZhang2023arXiv,Rubenstein2023arXiv} like text tokens in LLMs
or converted to token embeddings of LLMs by applying additional adapter modules
\citep{shu2023llasmlargelanguagespeech,Wu2023ASRU,Fathullah2024ICASSP},
which are parameterized by shallow neural networks,
and concatenated with text embeddings as inputs of LLMs.
Thanks to the contextualization of LLMs by prompting,
LLMs can perform not only automatic speech recognition (ASR) but also other tasks,
such as speech translation and voice chat \citep{Chu2024arXiv,Llama-Team-AI-Meta2024techreport}
by including instructions in the prompt texts to the LLMs.

Even to specialize ASR tasks, the LLM-based ASR systems can be contextualized
by providing prior information on input speech to the prompts
without modifying model architectures,
unlike conventional ASR contextualization techniques
\citep{Pundak2018SLT,Ding2019INTERSPEECH,Jain2020arXiv,Le2021INTERSPEECH}.
For example, Speech LLaMA \citep{Lakomkin2024ICASSP} improved ASR performance
by adding video titles and descriptions to the text prompts.
Although the Speech LLaMA provides more general information along with input speech,
providing individual words or phrases similarly to the conventional ASR contextualization techniques will enable LLM-based ASR systems to transcribe named entities or domain-specific terms.
Specifically, some languages have many homonyms, such as Japanese and Chinese (especially when dropping tones);
individual words or phrases provided as prior information
help LLMs to distinguish such homonyms.
Therefore, we develop an LLM-based ASR system that takes keywords as prior information.

The main contributions of this paper are that
(1) we develop LLM-based ASR systems
that take keywords as prior information in text prompts,
and (2) experimentally show that providing keywords as prior information
improves the recognition performance of rare and ambiguous words.

\section{Architecture}
\label{sec:architecture}

\subsection{Model Design}

\begin{figure}
      \centering
      \includegraphics[width=10cm]{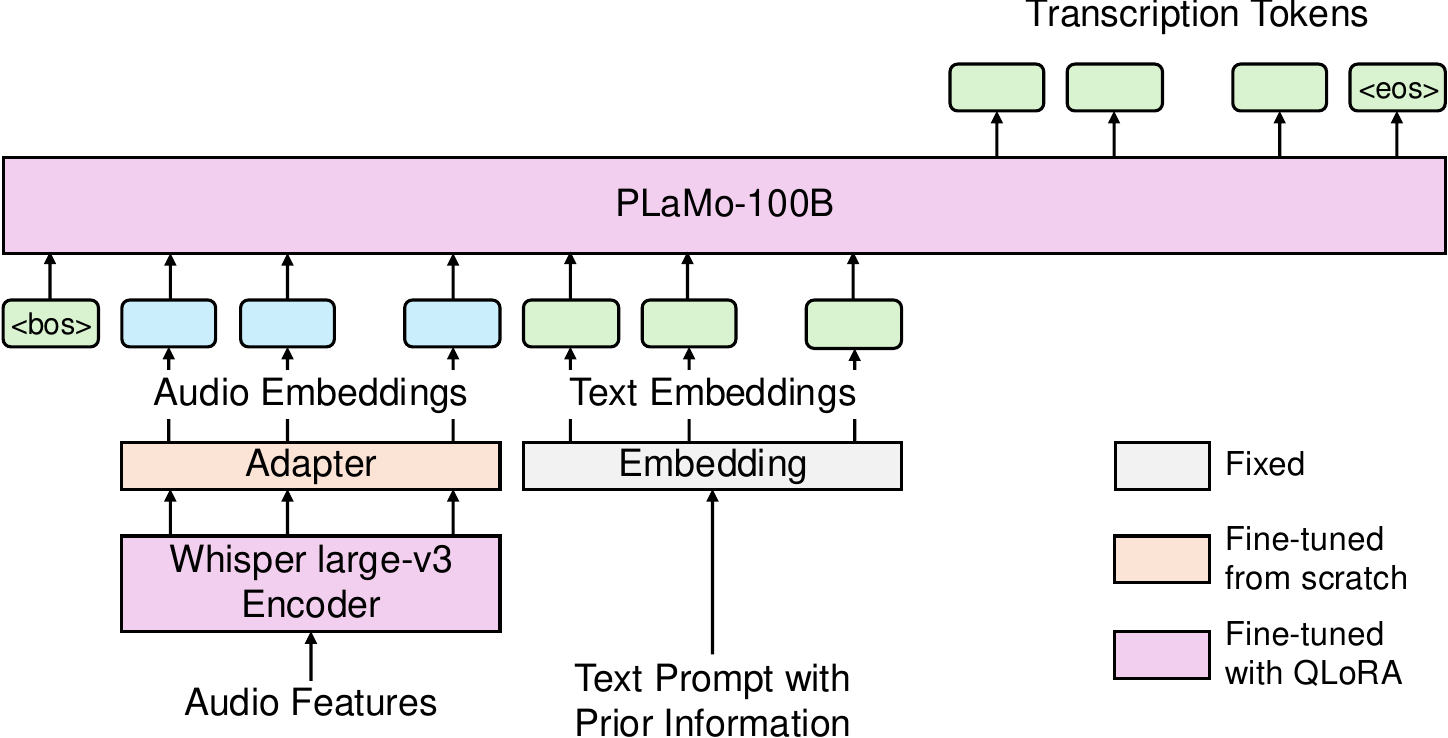}
      \caption{Model architecture}
      \label{fig:model-architecture}
\end{figure}

We adopt an in-house 100B-parameter LLM
pre-trained on Japanese and English datasets from scratch,
referred to as PLaMo-100B \citep{PFE2024arXiv},
as a decoder,
and the encoder of pre-trained Whisper large-v3%
\footnote{We used the checkpoint at \url{https://huggingface.co/openai/whisper-large-v3}.}
as an audio encoder, as shown in \Cref{fig:model-architecture}.
We also construct an LLM-based ASR system
using Swallow-7B \citep{fujii2024arXiv} as a decoder
that performed continual pre-training on LLaMA-2 7B \citep{Touvron2023arXiv}
on Japanese and English datasets, where Japanese was dominant.

In order to enjoy the expressiveness of
pre-trained encoder and decoder on massive data
and to simplify the training pipeline,
we do not modify the architecture of the pre-trained models and the input data format
as much as possible inspired by \citet{Lakomkin2024ICASSP}.
To do so, we feed projected audio features as input token embeddings of decoders.
Since the dimensionality of the original audio features extracted by the encoder
does not match with text embeddings in the decoder,
we introduce a linear adapter layer to project the original audio features to a text embeddings space.
We stack four consecutive audio features before applying the adapter
to reduce the length of audio tokens \citep{Fathullah2024ICASSP}.
The total number of parameters of the adapter is
$(4 \times \texttt{Audio feature dim}) \times \texttt{token dim}$ without bias.

Our 100B model does not fit in a single NVIDIA A/H100 GPU,
whose memory size is 80GB.
To overcome this limitation,
we use QLoRA fine-tuning \citep{Dettmers2023NeurIPS},
where the encoder and decoder's linear weights are quantized \citep{Dettmers2023NeurIPS}
and perform LoRA \citep{Hu2022ICLR} based fine-tuning.
Note that the linear adapter layer is not quantized
because this is initialized with random weights rather than pre-trained weights.

\subsection{Prompt Design}

To contextualize the decoders,
we feed additional prior information in the prompt \citep{Lakomkin2024ICASSP}.
We do not add special tokens to the decoder tokenizer like Whisper \citep{RadFord2022arXiv}
to fix pre-trained embedding layers of the decoders.
We design two prompt templates depending on the language: Japanese or English.

As a part of the prior information in the prompt,
we give a language identifier
inspired by Whisper's language tag \citep{RadFord2022arXiv}.
Concretely, we give simple texts
``\texttt{Language:~en}'' for English and
\begin{CJK}{UTF8}{ipxm}
      ``\texttt{言語：ja}'',
\end{CJK}
which has the same meaning in Japanese, for Japanese, respectively.

For further contextualization to transcribe rate and ambiguous words,
we give keywords of the transcriptions,
e.g., ``\texttt{Keywords:~tokyo, machine learning, speech, large language model}'',
following \citet{Lakomkin2024ICASSP}.
We use \texttt{NA} representing ``not available''
instead of keywords for samples not having keywords.
\Cref{sec:keyword-generation} describes how to create keywords.

In summary, our input to the decoder for English datasets%
\footnote{\Cref{sec:japanese-prompt} gives the prompt template for Japanese datasets.}
is as follows:

\begin{center}
      \small
      \texttt{<bos>[audio embeddings] Language:~en ; Keywords:~[keywords] ;
      Transcription:~[transcription]<eos>},
\end{center}

\noindent
where \texttt{<bos>} and \texttt{<eos>} are special tokens pre-defined by the decoder's tokenizer,
\texttt{[audio embeddings]} is a sequence of audio features
extracted by the audio encoder and the adapter,
\texttt{[keywords]} is a list of keywords as shown above,
and \texttt{[transcription]} is a ground-truth audio transcription.
At inference, we generate \texttt{[transcription]<eos>}
conditioned on its prefix prompt, i.e., audio embeddings and text prior.
Note that depending on the pre-training procedure of the decoder,
the \texttt{<eos>} token in the prompt is set to be identical to the \texttt{<bos>} token
to align the input format for packing \citep{Brown2020NeurIPS},
where multiple sentences are packed into a single sequence of tokens
and \texttt{<bos>} is the delimiter between two sentences.

\section{Datasets}

We combined four publicly available datasets to fine-tune our models.
Since our decoders and the audio encoder were pre-trained
on at least Japanese and English datasets,
we included an English dataset as one of four datasets
to avoid catastrophic forgetting in the English domain.
We evaluated ASR performance on four datasets covering Japanese and English.
\Cref{tab:training-dataset,tab:eval-datasets} summarize the training and evaluation datasets.
We briefly explain datasets in the following.

\paragraph{CommonVoice v8.0/16.1}
We used Japanese subset of CommonVoice v8.0/16.1 \citep{Ardila2020LREC}\footnote{\url{https://commonvoice.mozilla.org/en/datasets}}.
We used the original dev/test sets in CommonVoice v8.0 and v16.1 for evaluation.
In addition, we used the train set of CommonVoice v16.1 for training.

\paragraph{LibriSpeech}
As the English dataset,
we used LibriSpeech \citep{Panayotov2015ICASSP},
which is a commonly used benchmark dataset.
We used the original train/dev/test sets in our experiments.
Although dev/test sets are further divided into clean/other sets,
that is, dev-clean/dev-other and test-clean/test-other sets,
we will report experimental results for merged dev/test sets
rather than individual clean/other sets for simplicity.

\paragraph{ReazonSpeech}
We used a subset of ReazonSpeech v1~\citep{Yin2023NLP} for training,
where we removed samples not included in ReazonSpeech v2.
The ReazonSpeech dataset was automatically constructed
by bootstrapping labeling processes
using ASR models on Japanese television shows.
As mentioned in the limitation of this dataset \citep[Sec. 3]{Yin2023NLP},
the original transcriptions of the television shows are inaccurate.
In our experiments, we also faced a missing transcriptions issue,
and we will discuss a few approaches to mitigate this issue in \Cref{sec:prompt-bias}.

\paragraph{YODAS}
We used the Japanese subset of YODAS v1 \citep{Li2023ASRU},
which is constructed from YouTube videos with a Creative Commons license.
The subset has two further subsets
depending on how transcriptions are generated:
manual (\texttt{ja000}) or auto-generated (\texttt{ja100}).
Even in the manual subset,
some samples are not appropriate for ASR because their captions explain the whole video or background sounds.
Thus we selected samples whose quality is estimated to be relatively high
on a video-by-video basis.
Concretely, we filtered out videos
satisfying at least one of the following conditions:
1) character error rates (CERs) obtained by Whisper large-v3 of $40\%$ or higher and
2) ratios of alphabetic (\texttt{[a-zA-Z]}) characters of $50\%$ or higher.
Then from the manual subset,
we selected $10$ hours of high-quality videos
($87$ videos with $8\,313$ audio samples in total) as the test set
by checking audio samples and transcriptions manually,
and $10$ hours of videos randomly ($81$ videos with $8\,183$ audio samples in total)
except for videos in the test set as a dev set.
Finally, the complement set of the dev and test sets is the train set.
As a result, the test set is clearer than the train/dev sets,
and the dev set is clearer than the train set.

\begin{table}
      \centering
      \caption{Training dataset. In the training dataset, only YODAS is used twice. See \Cref{sec:prompt-bias} for the reason.}
      \label{tab:training-dataset}
      \begin{tabular}[t]{cccrr}
            \toprule
            Name & Language & Keywords & \# samples & Hours \\
            \midrule
            CommonVoice v16.1 & Japanese & & $9\,616$ & $13.0$ \\
            LibriSpeech & English & & $281\,241$ & $961.1$ \\
            ReazonSpeech & Japanese & & $11\,073\,520$ & $18\,867.5$\\
            YODAS & Japanese & $\checkmark$ & $1\,080\,953$  & $1\,137.3$ \\
            \midrule
            Total & & & $13\,703\,510 $ & $22\,713.6$ \\
            \bottomrule
      \end{tabular}
\end{table}

\begin{table}
      \centering
      \caption{Evaluation datasets}
      \label{tab:eval-datasets}
      \begin{tabular}[t]{ccccrr}
            \toprule
            Name & Set & Language  & Keywords & \# samples & Hours \\
            \midrule
            CommonVoice v8  &  dev & Japanese & & $4\,124$ & $5.7$ \\
            CommonVoice v8  & test & Japanese & & $4\,483$ & $6.5$ \\
            CommonVoice v16.1  &  dev & Japanese & & $6\,094$ & $7.9$ \\
            CommonVoice v16.1  & test & Japanese & & $6\,094$ & $8.9$ \\
            LibriSpeech &  dev &  English & & $5\,567$ & $10.5$ \\
            LibriSpeech & test &  English & & $5\,559$ & $10.7$ \\
            YODAS & dev  & Japanese & $\checkmark$ & $8\,183$ & $10.0$ \\
            YODAS & test & Japanese & $\checkmark$ & $8\,313$ & $10.0$ \\
            \bottomrule
      \end{tabular}
\end{table}

\subsection{Keyword Generation}
\label{sec:keyword-generation}

We generated keywords for the YODAS datasets using another pre-trained language model
because all the ASR datasets do not explicitly provide keywords for the transcriptions.
We extracted keywords for each video
and used them for all the pairs of audio samples and transcriptions derived from the same video
since a pair of audio and transcription is a short clip of the original video in the YODAS datasets.
Thus, not all keywords appear in their transcriptions.
We believe this is a practical usage of keywords;
a user defines keywords once for a single audio recording
and reuses the keywords for all audio clips extracted from the recording
rather than giving different keywords for every shorter clip.

To generate the keywords, we used Japanese Stable LM Base Gamma 7B%
\footnote{\url{https://huggingface.co/stabilityai/japanese-stablelm-base-gamma-7b}} model.
Generated keywords differed depending on random seeds used for the inferences of the model.
Due to this randomness, we extracted keywords
with different random seeds multiple times
and kept keywords generated $\nicefrac{1}{3}$ or more times.

\subsection{Dataset Bias via Prompt}
\label{sec:prompt-bias}

In our preliminary experiments,
we found performance degradation due to keyword conditioning on the dataset
because we only introduced the keywords to the YODAS dataset
rather than multiple datasets during fine-tuning.
We suspect that, as a result, the decoders can memorize the training dataset's bias
depending on whether keywords exist or not.
Concretely, the ReazonSpeech dataset,
which is the majority of the training dataset,
tends to drop a part of the transcription text
even if it exists in the speech part as the limitation in \citet{Li2023ASRU}.
Thus, if we give the placeholder keyword,
\begin{CJK}{UTF8}{ipxm}
      ``\texttt{キーワード：なし}'',
\end{CJK}
which is identical to ``\texttt{Keywords: na}'' in Japanese,
to the prompts for samples from the YODAS dataset at inference,
our fine-tuned models transcribe the audio as if a sample from ReazonSpeech;
it drops the transcription partially.
This missing transcription error increases CERs due to deletion errors
as demonstrated in \Cref{sec:bias-experiments}.
This tendency was more severe with the PLaMo-100B-based system than the Swallow-7B's counterpart in our preliminary experiments.

To mitigate this problem, we found approaches for fine-tuning and inference, respectively.
For the fine-tuning step,
we could reduce the bias by adding the YODAS dataset without keywords to the training dataset,
not to condition the placeholder keywords on the ReazonSpeech dataset.
For the inference step, we could reduce the degradation
by using random keywords to replace the placeholder keyword.
Unfortunately, the inference approach introduces additional latency,
though it mitigates the performance degradation.
Therefore, we used the approach for fine-tuning in our final experimental configuration.
As demonstrated in experiments,
we could only overcome this issue partially, which is future work.

\section{Experiments}
\label{sec:experiments}

We verified our ASR systems fine-tuned on the constructed training dataset
by conducting numerical experiments.
As a controlled experiment to evaluate keywords-based contextualization,
we also evaluated our ASR systems with slightly modified datasets and prompts;
we did not feed keywords even if they were available and removed keyword text from prompts.
We randomly shuffled keywords during fine-tuning and inference to make the model more robust regarding keyword ordering.

\subsection{Baseline Model}
We used pre-trained Whisper large-v3 downloaded from Hugging Face as the baseline system.
Whisper has a typical transformer encoder-decoder architecture.
It differs from our models in that
the encoder and the decoder are connected through cross-attention
instead of the adapter layer and input token embeddings.
We explicitly specified the Whisper's language tag for a fair comparison with our language identifier in the prompt.
We used \texttt{bf16} for its weights to fasten inference speed.

\subsection{Pre-processing and Post-processing of Texts}

\paragraph{Pre-processing}
We applied Whisper's text normalizers%
\footnote{\url{https://github.com/openai/whisper/tree/main/whisper/normalizers}}
to the transcriptions and keywords
to use consistent pre-processing with the Whisper baseline.
In addition, we removed unnecessary whitespaces between non-ASCII characters
for the Japanese datasets.
After tokenization, we truncated keyword tokens
so that the sum over lengths of prompt, keywords, and transcription
becomes less than or equal to $300$ for our ASR systems.

\paragraph{Post-processing}
We noticed that the pre-trained Whisper model transcribed
characters or symbols that the text normalizers should remove.
As a result, Whisper's evaluation metric values were severely worse than our systems.
For a fair comparison, we applied the text normalizers to the transcriptions generated by Whisper and our systems.
This post-processing mainly boosted the performance of the Whisper model because our systems rarely generated such unintended characters and symbols.
For example, CERs on the YODAS dev set of Whisper large-v3
were $16.88$ and $20.78$ with and without post-processing, respectively.

\subsection{Optimization}

We fine-tuned our models for only one epoch due to computational constraints.
We used AdamW optimizer \citep{Loshchilov2019ICLR}
with paged implementation by \texttt{bitsandbytes}%
\footnote{\url{https://github.com/TimDettmers/bitsandbytes}}
and gradient checkpointing to reduce peak GPU memory
to fine-tune PLaMo-100B-based models on even one H100 GPU.
We used a linear warmup with a cosine annealing learning scheduler \citep{Loshchilov2017ICLR},
whose warmup was $1\%$ iterations.
We scaled the base learning rate with square root scaling
depending on the number of samples in mini-batches \citep{Hoffer2017NeurIPS}:
$\mathtt{lr} = \mathtt{base\_lr} \times \sqrt{\mathtt{bs}}$,
where $\mathtt{bs}=\mathtt{mini\_batch\_size\_per\_gpu} \times \mathtt{num\_gpus}$.
We minimized the cross-entropy loss over transcription tokens.

\paragraph{PLaMo-100B hyper-parameters}
In AdamW's hyper-parameters,
we decreased $\beta_2=0.95$ from the default value $0.999$
by following \citet{Wortsman2023NeurIPS}
because we observed that fine-tuning PLaMo-100B-based models
is more likely to face loss spike issue%
\footnote{\Cref{sec:stability-attempts} shares other attempts to stabilize fine-tuning.}.
We used $\mathtt{base\_lr}=3.5 \times 10^{-6}$.
We used a mini-batch size of $6$ per GPU
for both with and without keyword settings
for pair comparison in terms of the number of gradient updates.
We accelerated the training using distributed data-parallel \citep{Li2020VLDB}
with $72$ H100 GPUs.

We used $r=12$ and $\alpha=2 \times r$ for QLoRA without dropout \citep{Srivastava2014JMLR}.
As a result, the number of learnable weights
is $218\,234\,880$ out of $103\,098\,001\,920$ ($0.21\%$).
Note that as an implementation trick of the PLaMo-100B architecture,
self-attention's three linear matrices for query/key/values \citep{Vaswani2017NeurIPS}
are parameterized as a single matrix,
and its transformed features are split into query, key, and value representations.
Not to modify this implementation,
we introduced LoRA to the single-matrix
rather than three query, key, and value matrices as in the native implementation.
As a result, the number of LoRA weights for a set of query, key, and value of PLaMo-100B
is $(3 d_k \times r + r \times d_\mathrm{model})$%
\footnote{The terms $d_k$ and $d_\mathrm{model}$ are the same as in \citet{Vaswani2017NeurIPS}.}
instead of $3 (d_k \times r + r \times d_\mathrm{model})$
for a single head attention with the same dimensionality of
independent key, query, and value representations.
For the quantization of base weights for LoRA,
we used double quantized with \texttt{nf4} \citep{Dettmers2023NeurIPS}
implemented by \texttt{bitsandbytes}.
During forward computation, quantized weights are de-quantized with the \texttt{bf16} format.

\paragraph{Swallow-7B hyper-parameters}
We fine-tuned \texttt{Swallow-7b-plus-hf}%
\footnote{\url{https://huggingface.co/tokyotech-llm/Swallow-7b-plus-hf}}
as a smaller size decoder than PLaMo-100B.
We only describe the different hyper-parameters from PLaMo-100B's setting.
We could use the default $\beta_2=0.999$ in AdamW's hyper-parameters without frequent loss spikes.
We used $\mathtt{base\_lr}=7.5 \times 10^{-6}$
and mini-batch size of $64$ per GPU.
We used $8$ GPUs for distributed data-parallel
to perform a similar number of gradient updates to PLaMo-100B's fine-tuning.

We independently introduced LoRA weights to query, key, and value matrices
because of the implementation difference from PLaMo-100B.
We used $r=16$ and $\alpha=2 \times r$ for QLoRA without dropout.
As a result, the number of learnable weights
is $72\,749\,056$ out of $7\,539\,687\,424$ ($0.96\%$).

\subsection{Transcription Generation}
\label{sec:generation-details}

We transcribed audio with prompts
using the greedy search for all baseline/fine-tuned models and datasets.
For YODAS dev/test sets,
we also transcribed audio without keywords
to show the difference at inference depending on the existence of keywords.
In our preliminary attempts,
we allowed all models to generate $444$ tokens for transcriptions,
which is the maximum number of tokens of Whisper large-v3 when specifying the language.
However, we observed that this value caused too many insertion errors
due to meaningless repetitions in several transcriptions,
unnecessarily worsening evaluation metrics.
To mitigate these undesirable insertions,
we computed the max token lengths of dev sets for each dataset and tokenizer,
as shown in \Cref{tab:dev-set_max_token_length} of \Cref{sec:details-experiments},
and used $1.25$ times those lengths as the max token lengths for transcription generation
for both dev and test sets.

\subsection{Evaluation Metrics}

As ASR performance metrics,
we will report character error rates (CERs) for the Japanese datasets: CommonVoice and YODAS,
and word error rates (WERs) for the English dataset: LibriSpeech.
We will also report keyword error rates (KWERs) for the YODAS dev/test sets,
where we selected keywords with a frequency of occurrence of $0$ in train sets for evaluation
from those generated for dev/test sets.

\subsection{Results}

\begin{table}[t]
      \centering
      \caption{
            Evaluation metrics comparison.
            Reported metric values are Character Error Rates (CERs)
            except for the LibriSpeech dataset, whose metric is Word Error Rates (WERs).
            ``KW'' stands for keywords in the header.
            In the second and third columns,
            check marks $\checkmark$ represent the existence of keywords
            at the training and inference steps, respectively.
            }
      \label{tab:cer-results}
      {\footnotesize
      \include{tables/metric}
      }
\end{table}

\begin{table}[t]
      \centering
      \caption{
            Keywords error rate comparison.
            ``KW'' stands for keywords in the header.
            In the second and third columns,
            check marks $\checkmark$ represent the existence of keywords
            at the training and inference steps, respectively.
            By using keywords for inference,
            the evaluation metric values become much lower than not using keywords.
      }
      \label{tab:kwer-results}
      \include{tables/kwer}
\end{table}

\Cref{tab:cer-results} shows CERs and WERs on dev/test sets.
CERs for Japanese datasets improved by replacing the Whisper decoder with PLaMo-100B or Swallow-7B,
probably because both models have a much larger number of parameters than the Whisper decoder,
and training datasets for pre-training and fine-tuning of the models
were dominated by Japanese text/audio data.
In addition, compared to the case without keywords,
CERs for YODAS dev/test sets improved significantly
by giving keywords for training and inference
with relative error rate reductions of
$7.87\%$ (from $12.45\%$ to $11.47\%$ in CERs) for the dev set
and $11.15\%$ (from $10.67\%$ to $9.48\%$ in CERs) for the test set for PLaMo-100B.
On the other hand, compared to Whisper large-v3,
the WERs of our models for LibriSpeech degraded
due to the small number of English samples in the training dataset,
though our models could avoid catastrophic forgetting in the English domain.

\Cref{tab:kwer-results} shows KWERs for YODAS dev/test sets.
As expected,
we could significantly improve KWERs by giving keywords as prior information.
Our model could not only improve transcriptions in terms of pronunciation
but also could select correct homonyms for named entities,
such as domain-specific terms and personal names,
by contextualizing the decoder with keywords
as shown in \Cref{tab:examples}.

Comparing the results of PLaMo-100B with those of Swallow-7B in \Cref{tab:cer-results},
although PLaMo-100B outperformed Swallow-7B
for most of the evaluation sets except for the YODAS dev set,
improvements in CERs and KWERs were smaller
than expected from the increase in model size.
One of the reasons is that
the number of training epochs was insufficient for PLaMo-100B
due to limitations on computational costs.
Another reason is that
speech recognition is easier than other tasks,
such as speech translation and voice chat,
and 7B models may be sufficient for speech recognition
in terms of model sizes.
Indeed, this tendency is consistent with
\citet[Table 31]{Llama-Team-AI-Meta2024techreport}
reporting that LLaMA 3.1 8B and 70B-based multi-task speech processing systems' WERs,
where their difference was only $1.4$ at maximum.
Our future work is fine-tuning our model with PLaMo-100B using more data and the number of epochs and extending to other tasks.

\newcommand*{\cjkm}[1]{\begin{CJK}{UTF8}{ipxm}#1\end{CJK}}
\newcommand*{\cjkg}[1]{\begin{CJK}{UTF8}{ipxg}#1\end{CJK}}
\begin{table}[t]
      \centering
      \caption{
            Examples of recognition results in Japanese from YODAS dev/test sets
            improved by providing keywords as prior information.
            Keywords and corresponding recognized words are shown in bold (gothic),
            and recognition errors are marked by underlines.
      }
      \label{tab:examples}
      \small
      \begin{CJK}{UTF8}{ipxm}
      \begin{tabular}[t]{p{0.3\textwidth}p{0.3\textwidth}p{0.3\textwidth}}
      \toprule
      Ground truth & Without keywords & With keywords \\
      \midrule
      \cjkg{阿蘇ローズベリー香園}\cjkm{と} &
      \cjkg{阿蘇ローズベリー\underline{公}園}\cjkm{と} &
      \cjkg{阿蘇ローズベリー香園}\cjkm{と} \\
      \midrule
      \cjkg{カラス天狗}\cjkm{のウルトラマン} &
      \cjkg{カ\underline{ー}ス\underline{ティング}}\cjkm{のウルトラマン} &
      \cjkg{カラス天狗}\cjkm{のウルトラマン} \\
      \midrule
      \cjkm{ラズパイで}\textbf{handbrake}\cjkm{いいですね} &
      \cjkm{ラズパイで}\cjkg{\underline{ハンドブレーキ}}\cjkm{いいですね} &
      \cjkm{ラズパイで}\textbf{handbrake}\cjkm{いいですね} \\
      \midrule
      \cjkg{堂満直聖}\cjkm{26歳です} &
      \cjkg{\underline{道}満直\underline{輝}}\cjkm{26歳です} &
      \cjkg{堂満直聖}\cjkm{26歳です} \\
      \midrule
      \cjkg{塚本秀春}\cjkm{さんが代表で鍵のレプリカを受け取り} &
      \cjkg{塚本\underline{英治}}\cjkm{さんが代表で鍵のレプリカを受け取り} &
      \cjkg{塚本秀春}\cjkm{さんが代表で鍵のレプリカを受け取り} \\
      \midrule
      \end{tabular}
      \end{CJK}
\end{table}

\subsection{Dataset Bias Experiments}
\label{sec:bias-experiments}

To demonstrate the issue discussed in \Cref{sec:prompt-bias},
we conducted a controlled experiment using Swallow-7B models.
We fine-tuned the same pre-trained Swallow-7B model,
represented by cross marks \xmark\ in \Cref{tab:bias-results},
on another training dataset that always feeds keywords for the YODAS train set;
samples with keywords in the YODAS train set were shown twice to the model in the single epoch.
\Cref{tab:bias-results} reports KWERs and CERs on the YODAS dev set.
KWERs for transcriptions with keywords
improved by using the model fine-tuned on the new train set represented by \xmark,
possibly by increasing the amount of training data with keywords.
However, KWERs and CERs were worst when keywords were not provided to the model at inference
(the second row from the bottom of \Cref{tab:bias-results}),
especially regarding deletions, which almost doubled.
As described in \Cref{sec:prompt-bias},
we suspect this is because only the YODAS train set has keywords,
and other samples, mainly ReazonSpeech, do not.
Thus, ReazonSpeech's transcription error,
which tends to drop a part of the transcription text,
hurt the transcription of even YODAS dev/test sets
by not giving keywords at inference.

\begin{table}[t]
      \centering
      \caption{
            Detailed error analysis on the YODAS dev set with Swallow-7B models.
            The cross marks \xmark\ represent the use of the YODSA train set with keywords only.
            Recall that check marks $\checkmark$ represent
            the use of the YODAS train set with and without keywords,
            and no marks represent the use of the set without keywords only.
            The amount of data of the train set is the same for all conditions;
            each audio sample in the YODAS train set is shown to the model twice in a single epoch.
      }
      \label{tab:bias-results}
      \include{tables/yodas_dev_swallow_errors}
\end{table}

Recall that
we added both the YODAS dataset without and with keywords
to mitigate the issue of biasing to ReazonSpeech
for the training dataset with keywords for fine-tuning, which is represented by check marks $\checkmark$ at ``KW@Train'' in
\Cref{tab:cer-results,tab:kwer-results,tab:bias-results}.
However, comparing CERs without keywords of PLaMo-100B models in \Cref{tab:cer-results},
the model fine-tuned without keywords outperformed the model fine-tuned with keywords
on CommonVoice v16.1, LibriSpeech, and YODAS datasets.
We suspect that the size of ReazonSpeech is still much larger than other datasets,
and adding YODAS without keywords was not sufficient to completely solve this issue,
resulting in performance degradation of fine-tuning with keywords
compared to fine-tuning without keywords,
where YODAS without keywords was used twice in a single epoch
to make the train data size equivalent.
We suggest diversifying the training data as much as possible
if the goal is to train generalized ASR systems that can take prior texts via prompts.

\section{Conclusion}

We constructed an ASR system based on an in-house LLM PLaMo-100B
combined with the Whisper large-v3 encoder.
By contextualizing the LLM-based decoder
using keywords directly as prior information in the text prompt,
we showed that recognition performance for keywords
that did not appear in the train set
improved significantly in correcting pronunciations and selecting homonyms.
In future work,
training our models using more data, including other tasks and languages, is a promising direction to bring out the potential of our pre-trained LLMs.

\subsubsection*{Author Contributions}

Kento Nozawa implemented the training and evaluation pipeline,
conducted numerical experiments,
and drafted the manuscript.
Takashi Masuko constructed all datasets and extracted their keywords,
implemented keywords error evaluation,
and wrote the manuscript.
Toru Taniguchi prototyped and supported the training and evaluation pipeline,
revised the manuscript,
and managed this project.

\subsubsection*{Acknowledgments}

This paper is based on results obtained from the project, ``Research and Development Project of the Enhanced Infrastructures for Post 5G Information and Communication System'' (JPNP 20017), subsidized by the New Energy and Industrial Technology Development Organization (NEDO).

We would like to thank (ex-) PFN/PFE members, especially Daiki Higurashi, Linsho Kaku, Toshiki Kataoka, Hiroaki Mikami, Shintarou Okada, Daisuke Okanohara, Shuji Suzuki, Daisuke Tanaka, Seiya Tokui, Zijian Xu, and Shoichiro Yamaguchi, for their helpful discussions and/or implementation support; Toshihiko Yanase, Yuta Hirokawa, and Toru Komatsu for infrastructure-related support.

\bibliography{reference}
\bibliographystyle{plainnat}

\appendix

\section{Japanese Prompt}
\label{sec:japanese-prompt}

We used the following prompt for Japanese datasets:

\begin{CJK}{UTF8}{ipxm}
      \begin{center}
            \small
            \texttt{<bos>[audio embeddings] 言語：ja； キーワード：[keywords]； 書き起こし：~[transcription]<eos>},
      \end{center}

      where an example of keywords is ``\texttt{東京、機械学習}'' and use ``\texttt{なし}'' if keywords are unavailable.
\end{CJK}

\section{Details of Experiments}
\label{sec:details-experiments}

\Cref{tab:dev-set_max_token_length} shows the max token length of dev sets using corresponding tokenizers.
Note that the numbers included the special token's length, specifically, the single end of the token.

\begin{table}
      \centering
      \caption{
            Max token lengths of transcriptions for dev sets.
            }
      \label{tab:dev-set_max_token_length}
      \include{tables/dev_set_token_length}
\end{table}

\subsection{Attempts to Stabilize Fine-tuning}
\label{sec:stability-attempts}

In our preliminary experiments, we sometimes encountered loss spikes during the fine-tuning that caused severe ASR performance degradation.
We share our attempts to resolve these loss spikes below.

\paragraph{Mini-batch creation}

We attempted to accelerate the training speed by using length-grouped mini-batches implemented by \texttt{transformer}%
\footnote{\url{https://github.com/huggingface/transformers/blob/v4.43.3/src/transformers/trainer\_pt\_utils.py\#L593-L625}}
to make sure that each mini-batch contains similar sequence lengths.
However, this causes training loss spikes
especially when the length of the sequence largely changes across mini-batches.
By following \citet{Morishita2017NMTWorkshop}
that conducted similar experiments on neural machine translation with non-transformer models,
we used random shuffling to create mini-batches instead of length grouping.

\paragraph{Optimizer's hyper-parameters}
Even though we used randomly shuffled mini-batches,
the loss spike made the fine-tuning of PLaMo-100B models unstable.
Similar to experimental configurations
in \citet{Touvron2023arXiv} and \citet{Wortsman2023NeurIPS},
reducing AdamW's $\beta_2$ was effective
in avoiding the loss spikes to fine-tune PLaMo-100B models.

\paragraph{Auxiliary loss}
By following the pre-training of PLaMo-100B,
we also introduced the z-loss \citep{Chowdhery2023JMLR} with the coefficient term of $1e^{-4}$
as an auxiliary loss to stabilize fine-tuning.
However, loss spikes still happened with AdamW,
whose $\beta_2=0.999$ for PLaMo-100B,
and reducing $\beta_2$ was helpful enough to avoid the spikes.
As a result, we excluded the z-loss to avoid additional regularization.

\subsection{Libraries Used in Implementation}

We could conduct our experiments smoothly thanks to the following open-sourced software.
All experiments were tracked by using \texttt{wandb} \citep{wandb} that GENIAC provided.
In our implementations, we used \texttt{transformers} \citep{Wolf2020ACLDemo} for encoder and decoder models' implementations.
As LoRA's implementation, we used \texttt{peft} \citep{peft}.
In addition, we used \texttt{PyTorch} \citep{Ansel2024ASPLOS} for the training and evaluation pipeline and \texttt{torchaudio} \citep{Hwang2023torchaudio} for data-related implementation.
We created our \LaTeX\ tables by using \texttt{pandas} \citep{reback2020pandas}.

\end{document}

%% file: math_commands.tex
\usepackage{amsmath,amsfonts,bm}

\def\eqref#1{equation~\ref{#1}}

\def\1{\bm{1}}

\DeclareMathAlphabet{\mathsfit}{\encodingdefault}{\sfdefault}{m}{sl}
\SetMathAlphabet{\mathsfit}{bold}{\encodingdefault}{\sfdefault}{bx}{n}



%% file: tables/metric.tex
\begin{tabular}{lllrrrrrrrr}
\toprule
Model & KW@Train & KW@Inference & \multicolumn{2}{r}{CommonVoice v8.0} & \multicolumn{2}{r}{CommonVoice v16.1} & \multicolumn{2}{r}{LibriSpeech} & \multicolumn{2}{r}{YODAS} \\
 &  &  & Dev & Test & Dev & Test & Dev & Test & Dev & Test \\
\midrule
PLaMo &    &    & $4.37$ & $5.83$ & $\mathbf{9.57}$ & $\mathbf{13.30}$ & $3.47$ & $3.63$ & $12.45$ & $10.67$ \\
PLaMo & $\checkmark$ &    & $\mathbf{4.24}$ & $\mathbf{5.63}$ & $9.70$ & $13.67$ & $4.16$ & $4.38$ & $12.93$ & $11.29$ \\
PLaMo & $\checkmark$ & $\checkmark$ &     &     &     &     &     &     & $11.47$ & $\mathbf{9.48}$ \\
Swallow &    &    & $4.99$ & $6.77$ & $10.22$ & $13.98$ & $4.22$ & $4.04$ & $12.72$ & $11.18$ \\
Swallow & $\checkmark$ &    & $5.05$ & $7.24$ & $10.32$ & $13.85$ & $4.39$ & $4.41$ & $13.05$ & $11.06$ \\
Swallow & $\checkmark$ & $\checkmark$ &     &     &     &     &     &     & $\mathbf{11.37}$ & $9.54$ \\
Whisper &    &    & $6.60$ & $8.55$ & $11.92$ & $15.24$ & $\mathbf{2.89}$ & $\mathbf{2.95}$ & $16.88$ & $13.50$ \\
\bottomrule
\end{tabular}

%% file: tables/kwer.tex
\begin{tabular}{lllrr}
\toprule
Model & KW@Train & KW@Inference & \multicolumn{2}{r}{YODAS} \\
 &  &  & Dev & Test \\
\midrule
PLaMo &    &    & $63.77$ & $52.74$ \\
PLaMo & $\checkmark$ &    & $64.17$ & $58.21$ \\
PLaMo & $\checkmark$ & $\checkmark$ & $24.29$ & $\mathbf{10.45}$ \\
Swallow &    &    & $64.17$ & $55.47$ \\
Swallow & $\checkmark$ &    & $65.99$ & $57.71$ \\
Swallow & $\checkmark$ & $\checkmark$ & $\mathbf{21.46}$ & $10.95$ \\
Whisper &    &    & $71.66$ & $61.69$ \\
\bottomrule
\end{tabular}

%% file: tables/yodas_dev_swallow_errors.tex
\begin{tabular}{llrrrrr}
\toprule
KW@Train & KW@Inference & KWER & CER & Insertions & Deletions & Substitutions \\
\midrule
&    & $64.17$ & $12.72$ & $6\,105$ & $6\,961$ & $10\,519$ \\
$\checkmark$ &    & $65.99$ & $13.05$ & $6\,428$ & $6\,932$ & $10\,828$ \\
$\checkmark$ & $\checkmark$ & $21.46$ & $11.37$ & $5\,991$ & $6\,068$ & $9\,014$ \\
\xmark &    & $72.47$ & $16.69$ & $64\,32$ & $12\,621$ & $11\,886$ \\
\xmark & $\checkmark$ & $18.62$ & $11.65$ & $6\,556$ & $6\,037$ & $8\,997$ \\
\bottomrule
\end{tabular}

%% file: tables/dev_set_token_length.tex
\begin{tabular}{lrrrr}
    \toprule
    Model & CommonVoice v8.0 & CommonVoice v16.1 & LibriSpeech & YODAS \\
    \midrule
    PLaMo   & $51$ & $54$ & $122$ & $61$ \\
    Swallow & $52$ & $54$ & $132$ & $62$ \\
    Whisper & $60$ & $69$ & $117$ & $71$ \\
    \bottomrule
\end{tabular}